\documentstyle[12pt]{article}

\begin{document}

\begin{center}

{\Huge Statistical limits to Equity}\\
\vspace{12pt}

{\bf \small Daniel Badagnani}\\
\vspace{12pt}

{\it \bf abstract}\\

{\it We derive the most probable distribution of resources for a simple
society. We find that a probabilistic analysis forbids both too much and too
less equity, and selects instead a minimally ordered state. We give the
detailed calculations for a special model where the population and
resources are fixed, and resources are owned only by individuals. We show
that in general the equity is greater whenever the volume of the
indifference manifold grows faster as a function of individual rent.}

\end{center}

\section{Introduction}

Maybe the central task of the whole economical science is the clever
administration of the resources of a society in order for everyone to
fullfill their material needs. It is thus interesting the question of to
what extent one could, in principle, distribute richness with equity. At
first sight, it seems perfectly conceivable (altough higly uthopic) a
society in wich everyone access exactly the same 'ammount of richness'. Is
there some conceivable bound to get such a setting? Well, suppose we begin
distributting goods to N people: there is just one
'configuration' in wich everyone receives exactly the same ammount of any
good. If one decide being unfair by beneffiting just someone, there are N
ways to do so (one for every beneffited person). One can see that as we
increase the 'unfairness', the number of possible configurations grow
fastly. Thus, if we 'prepare the system' to be fair at the begining, by
exchange it will disorder with litle probability of reordering. This
reasoning puts a bound also on the unfairness: a few too welth people and
the rest with nothing is again an ordered state.

Here we will give the detailed mathematics for a 'dilute' society, for wich
the following postulates holds:

\begin{enumerate}

  \item{The population N remains fixed}

  \item{There is a fixed ammount of richness, R}

  \item{There are only individual patrimonia}

\end{enumerate}

Under these simplifications, we will find that we can find the probability
distribution for individual wealthiness, giving a precise meaning to the
above discussion. We will find the number of states for detailed
distributions giving rise to the same global state (overall 'unfairness').
Those who know statistical mechanics will find a very close resemblance with
ideal gases. Notice, however, that the following is not an analogy:
everything is derived under the given and a few more assumptions. That much
of the derivations are formally identical is quite amusing, but not that
surprising once one realizes that both are statistics on the ways one can
distribute something indestructible among a fixed number of 'objects'.

The conclussions drawn from this model are quite limited: richness is
created and destroyed, populations change and in complex societies large
ammounts of resourses belongs to 'composite objects' like associations,
corporations or even the whole society ('large range interactions'). Even
more seriously, actual societies are away from statistical equilibrium:
there is a typical 'relaxation time' needed for the society to exchange
goods untill it gets to an equilibrium, but growth (especially sectorial
growth) ussually happens in shorter periods, so equilibrium is not reached.
We know from other disciplines that systems far away from equilibrium can
get extremely striking complexity (life being a paradigmatic thermodynamical
example). Anyway, analyzing what happens at equilibrium is the mandatory
begining, and very important conclussions are available at this primitive
stage.

\section{Partition of resourses: the distribution function}

For clarity, we will suppose that each individual can have any patrimonium
from a discrete set ${r_1,r_2,...,r_k}$. Every $r_i$ could be composed of
many combinations of different goods, say there are $g_i$ such combinations
for each $r_i$. We also suppose that any 'microscopical state' (any specific
combination of goods) is accesible with equal probability for every
individual ('fair rules'). We will call $n_i$ to the number of people owning
$r_i$ each. Recall that the total population is $N$ and the total
wealthiness is $R$. We have the bounds

\begin{equation}
N=n_1+n_2+...+n_k
\label{TotalPopulation}
\end{equation}

\begin{equation}
R=n_1 r_1+n_2 r_2+...+n_k r_k
\label{TotalRichness}
\end{equation}

We will count the total number of configurations and find which is the most
probable distribution (that is, the $n(r)$ having the maximum amount of
states). First, notice that we can fill the state $r_1$ with one person in
$N g_i$ ways, then we have $N-1$ remaining persons to accomodate. Once $n_i$
persons have been placed into an $r_i$ we must remember to divide by the
$n_i!$ ways one does so, and we are left with

\begin{equation}
\Omega=\frac{N!g_1^{n_1}...g_k^{n_k}}{n_1!...n_k!}
\label{NumberOfStates}
\end{equation}

We now suppose that there are lots of $n_i$'s at each $r_i$ level, and we
take them to be continuous. For each $r_i$ we will find an $n_i$ (we will
call it $n(r_i)$) wich maximizes \ref{NumberOfStates} subject to the bounds
\ref{TotalPopulation} and \ref{TotalRichness}. In order to do so, we
introduce the bounds as Lagrange multipliers, and extremize

\begin{equation}
S=ln(\Omega)+\alpha(N-n_1-n_2-...-n_k)+\beta(R-n_1 r_1-n_2 r_2-...-n_k r_k)
\label{extremize}
\end{equation}
In $ln(\Omega)$ we use Stirling formula. We get then

\begin{equation}
\frac{\partial S}{\partial n_i} = ln(g_i)-ln(n_i)-\alpha-\beta r_i=0
\label{Maximum}
\end{equation}
while the conditions $\frac{\partial S}{\partial \alpha}$ and
$\frac{\partial S}{\partial \beta}$ just imposes the bounds. Thus we have
got a Boltzman-like distribution for richness

\begin{equation}
n(r)=g_i e^{-\alpha}e^{-\beta r}
\end{equation}
with $\alpha$ and $\beta$ determined by the conditions
$$N=e^{-\alpha}\sum_r e^{-\beta r}$$
$$R=e^{-\alpha}\sum_r r e^{-\beta r}$$

\section{Particular models}

Lets consider some concrete distributions in order to see what do we have
here. The above distribution will be considered continuous, that is, n(r)
means that there are $n(r)dr$ individuals with patrimonia between $r$ and
$r+dr$. Then, since the probability is exponentially damped, we will forget
to bound r and we will take it to range between zero and infinity (in fact,
it couldnt exeed $R$, but the probability of being $R$ while one expect it
to be about $R/N$ is about $e^{-N}$, that is, astronomically small). We do
this because it is much easier to calculate. The distribution is thus

$$
n(r)=g(r)e^{-\alpha}e^{-\beta r}
$$

In order to give a $g(r)$ we will consider that we have $m$ goods whose
unitary price is much cheaper that the mean wealthy (we will clarify this
somewhat dark condition below) so we can think that we have a continuum of
goods. For a given $r$ we have an indifference hipersurface such that
$r=r_1+...+r_m$ ($r_i$ being the price of the ammount of the i-th good owned
by the individual). So, g(r) is proportional to the (m-1)-dimmensional
volume of the hypersurface: $S(r)=\sqrt{m} r^{m-1}$. That surface is an
hyperplane bounded by the walls defined by
$r_i=0$, and its 'volume' is of the form $Cr^{m-1}$. So we take as our
distribution

\begin{equation}
n(r)=Cr^{m-1}e^{-\alpha}e^{-\beta r}
\end{equation}

Let us first solve for the parameters $\alpha$ and $\beta$. We have

$$
N=C e^{-\alpha}\int_0^\infty r^{m-1}e^{-\beta r} dr
$$

$$
R=C e^{-\alpha}\int_0^\infty r^{m} e^{-\beta r} dr
$$
This gives the result

\begin{equation}
n(r)=\frac{N}{(m-1)!}\left(\frac{m}{\bar{r}}\right)^m r^{m-1}e^{-mNr/R}
\label{distributionM}
\end{equation}
where $\bar{r}=R/N$ is the mean rent per capita.
From \ref{distributionM} we can tell what is the condition for low unitary
prices: we mean that every price should be much lower than $\frac{R}{mN}$.
If one has many prices independent of $R$, as $R$ grows we should include
more and more goods, so $m$ depends indeed on $R$. We see that the more
goods available, the more peaky will the distribution be. As $m$ grows, the
position of the peak approach from below the rent per capita $R/N$. The
variance for the individual rent (that is rougly the dispersion of income)
is

\begin{equation}
\Delta r=\frac{1}{\sqrt{m}}\frac{R}{N}
\label{variance}
\end{equation}

\section{General Discussion}

For computational ease, we solved for an artificially simple example in wich
any good could be obtained from any combination summing the same price. In
general, however, one should have an 'indifference manifold' whose volume
will in general depend on the income. What is important for the calculations
above is the 'dispersion relation' connecting the individual rent and the
volume of the indifference manifold. If it follows a power law $V=Cr^m$, all
the above applies includding the calculation for the dispersion. It shows
that equity will grow if the elasticity $m$ grows.

We have shown that a closed economy at equilibrium cannot be perfectly fair.
Anyway, a complex society with large individual rent and large diversity can
reach a state quite close to equity. On the other hand, subsistence
economies in wich only a few goods are exchanged are mostly compossed of
poor individuals.

An important remark on fluctuations is in order here: the 'ocupation
density' $d=n(r)/N$ will indeed fluctuate around the distribution found, and
one should wonder how often it will differ appreciably from that
distribution. If the answer is 'very often', then all the analysis done is
nearly useless. The answer is found by calculating the standard deviation
$\Delta d$, which is found to be $d/\sqrt{N}$. One could be tempted to take
$N=6000000000$, but it must be remembered that larger populations have
larger time relaxations, depending on the intensity of the trade exchange.
That time for an entire continent would probably be many years, so normal
succeses impeed reaching equilibrium. Where interchange is strong enough in
order to have a somewhat small relaxation time
is inside a country, a city, etc where population is tipically a few million 
people, so fluctuations in $d$ can be estimated in $10^{-3}$ of $d$, wich is 
quite small. However, economies are continuously changing, and are surely 
much away from equilibrium. 

\end{document}